\documentclass[]{spie}  
\usepackage[]{graphicx}
\newcommand{\hthree}{$^3$He }
\newcommand{\hfour}{$^4$He }

\title{MAKO: A pathfinder instrument for on-sky demonstration of low-cost 350 micron imaging arrays}

\author{Loren J. Swenson\supit{a,b}, Peter K. Day\supit{b}, Charles D. Dowell\supit{b}, Byeong H. Eom\supit{a}, Matthew I. Hollister\supit{a,b}, Robert Jarnot\supit{b}, Attila Kov\'{a}cs\supit{a}, Henry G. Leduc\supit{b}, Christopher M. McKenney\supit{a}, Ryan Monroe\supit{b}, Tony Mroczkowski\supit{a,b}, Hien T. Nguyen\supit{b}, and Jonas Zmuidzinas\supit{a,b}
\skiplinehalf
\supit{a}California Institute of Technology, 1200 East California Blvd, Pasadena, California, United States; \\
\supit{b}NASA Jet Propulsion Laboratory, California Institute of Technology, 4800 Oak Grove Drive, Pasadena, California, United States
}
\authorinfo{Send correspondence to L.J.S.: E-mail: swenson@astro.caltech.edu}

\begin{document}
\maketitle

\begin{abstract}
Submillimeter cameras now have up to $10^4$ pixels (SCUBA 2). The proposed CCAT 25-meter submillimeter telescope will feature a 1 degree field-of-view. Populating the focal plane at 350 microns would require more than $10^6$ photon-noise limited pixels. To ultimately achieve this scaling, simple detectors and high-density multiplexing are essential. We are addressing this long-term challenge through the development of frequency-multiplexed superconducting microresonator detector arrays. These arrays use lumped-element, direct-absorption resonators patterned from titanium nitride films. We will discuss our progress toward constructing a scalable 350 micron pathfinder instrument focusing on fabrication simplicity, multiplexing density, and ultimately a low per-pixel cost.
\end{abstract}


\section{INTRODUCTION}
\label{sec:intro}  


Modern cryogenic detector arrays for infrared and submillimeter astronomy are now routinely achieving the kilopixel scale.\cite{hilton:513, niemack:690}  While this represents a significant technological milestone, there remains a strong demand for even larger arrays.  For example, the proposed CCAT 25-meter submillimeter telescope\cite{Astro2010} would require more than $10^6$ photon-noise limited pixels in order to fully sample the focal plane at a wavelength of $\lambda=350~\mu$m.  In order to achieve this considerable scaling, candidate cryogenic detector technologies must be highly multiplexed, simple to fabricate, and the per-pixel cost must be minimized.  One particularly attractive technology for achieving very large arrays is the superconducting microresonator detector.\cite{Day:2003, Mazin:2004,mazin:2009,Baselmans:2012,Zmuidzinas:2012}  The pixel in this case consists of a radio-frequency resonator patterned from a superconducting thin film. Incident energy absorbed by the resonator breaks superconducting Cooper pairs generating quasiparticles.  This perturbs the complex conductivity of the thin film $\delta \sigma(\omega) = \delta \sigma_1 - j \delta \sigma_2$, consequently producing a measurable shift in the frequency $\omega_r$ and quality factor $Q_r$ of the resonance.  Strong frequency-domain multiplexing is achieved by coupling an entire array of pixels to a radio-frequency transmission circuit; the frequency of each pixel is tuned by adjusting the resonator reactance.

\begin{figure}
\begin{center}
\begin{tabular}{c}
\includegraphics[width=.9\linewidth]{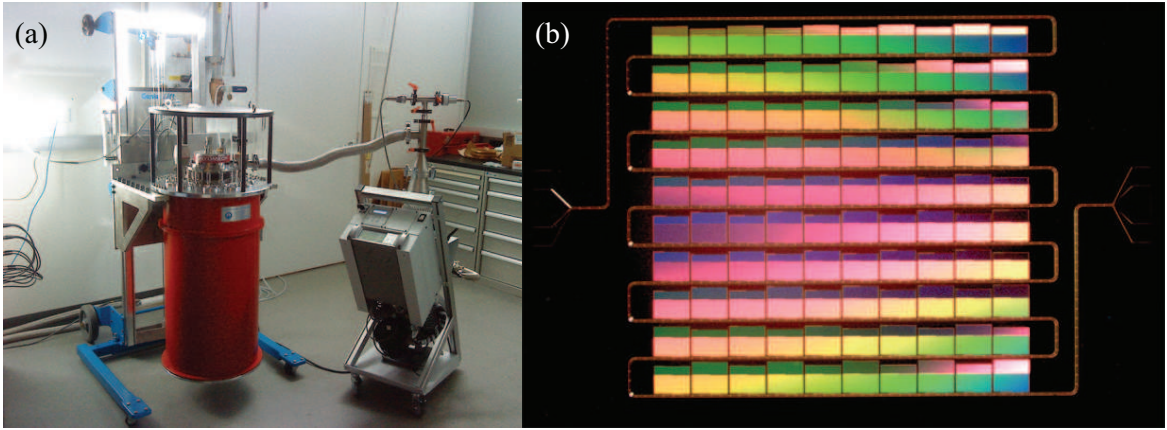}
\end{tabular}
\end{center}
\caption[example]
{
The MAKO instrument.  (a) \hthree measurement cryostat with a stable base temperature of 250 mK.  (b) First generation 100 pixel test array.  Here the single transimssion line is coupled to the inductive meander portion of each pixel and is used to read out the entire array.  For this initial design the capacitor occupies $\sim$ 35\% of the pixel area.}
\label{figSystem}
\end{figure}

Two methods are typically employed to couple optical signals into superconducting microresonator pixels.  One approach is to collect the incident radiation with an antenna and deposit the energy in the high-current portion of a quarter-wave coplanar waveguide resonator.\cite{schlaerth:2008, schlaerth:2012, monfardini:24, Lankwarden:2012}  This method has a number of advantages including good beam control, frequency selectivity, and the possibility of multi-chroic operation through the use of on-chip bandpass filters.  The second method utilizes a lumped-element resonator consisting of an inductive meander section with an attached interdigitated capacitor and is commonly referred to as the lumped-element kinetic inductance detector (LEKID).\cite{Doyle:2007,Doyle:2008}  In this design, radiation is directly absorbed by the inductive meander which is carefully designed to provide an effective far-infrared surface resistance of around $R_\mathrm{eff} = 377\,\Omega/(1 + \sqrt{\epsilon_r}) \approx 86\,\Omega$ when back-side illuminating the pixel through a silicon substrate with dielectric constant $\epsilon_r \approx 11.5$.  This result is modified to $R_\mathrm{eff}= 377 \, \Omega / \sqrt{\epsilon_r}$ if a backshort is used.  An important advantage of the lumped-element approach is that radiation absorbed by the meander is uniformly distributed over the inductor volume.  This avoids regions of high quasiparticle density which reduces the quasiparticle lifetime and consequently the device responsivity.  Lumped-element designs are also typically less complex to fabricate than antenna-coupled pixels, often requiring at most a few layers of lithography to realize an entire array.   On the other hand, good baffling and stray light rejection is relatively more important for absorber-coupled detectors.

For the LEKID, achieving a far-infrared surface resistance of $R_\mathrm{eff} \approx 86\,\Omega$ constrains the sheet resistance $R_s$ and area filling factor $\eta_A$ of the superconducting film by $R_\mathrm{eff} \approx R_s / \eta_A$.  To present a uniform absorber to incident radiation of wavelength $\lambda$, the meander spacing is further constrained by $s \ll \lambda$.  For $\lambda =$ 2 mm, near background-limited performance has been demonstrated using aluminum LEKIDs for ground-based astronomy.\cite{monfardini:24}  In this case, a low filling fraction $\eta_A$ was used to compensate for the low resistivity of aluminum.  At shorter wavelengths, the required meander spacing precludes the use of traditional elemental superconductors for reasonable film thicknesses and linewidths.  Fortunately, the discovery by Leduc et al.\cite{Leduc:2010} of the exceptional surface quality $Q_\sigma$ of superconducting titanium nitride (TiN) provides an alternative.  In this case, the resistivity $\rho_{TiN} \sim 100$ $\mu \Omega/$cm is several orders of magnitude higher than standard metals, allowing high fill fractions to be used with reasonable film thickness and optically lithographed linewidths.  Complete LEKID design details for the submillimeter/far-infrared are presented in an article in these proceedings by McKenney et al.\cite{mckenney:2012}

We are now actively engaged in constructing MAKO, a scalable 350 micron pathfinder instrument emphasizing multiplexing density, fabrication simplicity, and a low per-pixel cost.  An image of the cryostat and a 100 pixel first-generation test array are shown in Fig.\ \ref{figSystem}.  We have chosen LEKIDs patterned from TiN films for this project as the most suitable detector for achieving the project goals in the far-infrared.  Notably, these pixels are designed to resonate at $\approx 100$ MHz frequencies. All previous kinetic-inductance based imaging arrays have operated at $> 1$ GHz.  This shift toward lower frequencies provides numerous advantages including an improved pixel noise equivalent power (NEP), a simplified analog readout circuit, and a higher achievable multiplexing density.  These low frequency advantages are discussed below along with a detailed description of the cryo-mechanical and readout systems currently being commissioned.

\pagebreak

\section{Low frequency advantage}

The NEP of a superconducting microresonator detector is given by:
\begin{equation}\label{dNEP}
    \mathrm{NEP}_{diss}^2 = 2 P_0 h \nu(1+n_0) + NEP_{g-r}^2 + \frac{8 N_{qp}^2 \Delta_0^2}{\beta^2 \eta_0^2 \chi_c \chi_{qp}^2 \tau_{qp}^2}\frac{kT_a}{P_a}
\end{equation}
\begin{equation}\label{fNEP}
    \mathrm{NEP}_{freq}^2 = 2 P_0 h \nu(1+n_0) + NEP_{g-r}^2 + \frac{8 N_{qp}^2 \Delta_0^2}{\beta^2 \eta_0^2 \chi_c \chi_{qp}^2 \tau_{qp}^2}\frac{kT_a}{P_a} + \frac{8 N_{qp}^2 \Delta_0^2 Q_i^2}{\beta^2 \eta_0^2 \chi_{qp}^2 \tau_{qp}^2}S_{TLS}
\end{equation}
where Eq.\ (\ref{dNEP}) and Eq.\ (\ref{fNEP}) are the NEP when measuring perturbations of the resonator dissipation ($Q_r^{-1}$) and frequency, respectively.  In both of these equations the first three terms are the photon, generation-recombination, and amplifier noise contributions.  The full expression for the generation-recombination noise can be found in a recent review by Zmuidzinas\cite{Zmuidzinas:2012} but is not a significant noise source for the current designs where $h \nu \gg \Delta$.

The additional term in Eq.\ (\ref{fNEP}) reflects frequency noise due to dielectric fluctuations in the resonator capacitor.  Commonly referred to as two-level system (TLS) noise, these fluctuations arise from surface oxides or other amorphous dielectric material on the surface of the substrate.\cite{Gao:2008a,Gao:2008b}  TLS noise has been extensively studied and a variety of techniques have been developed to mitigate this effect.\cite{OConnell:2008,Noroozian:2009,Barends:2010}  One of the simplest methods for reducing TLS noise while simultaneously reducing the amplifier noise contribution is to increase the readout power.  For the simplest readout protocols, readout power is limited, this is limited by the onset of nonlinear effects and the eventual bifurcation of the device response.  Bifurcation occurs when
\begin{equation}\label{pgCritical}
    P_g^{(c)} = \frac{a_c Q_c \omega_r E_*}{2 Q_r^3}
\end{equation}
where $P_g^{(c)}$ is the generator power at bifurcation, $Q_c$ is the resonator coupling quality factor, $a_c = .77$ is the nonlinearity parameter at bifurcation, and $E_*$ sets the scaling of the nonlinearity.\cite{swenson:2012}  If the bifurcation mechanism is due to the nonlinearity of the kinetic inductance, $E_*$ is related to the condensation energy of the inductor $E_{cond} = N_0 V_L \Delta^2/2$, where $N_0$ is the material-dependent single-spin density of states at the Fermi energy, $\Delta \approx 3.5 k_B T_c /2$ is the superconducting gap, and $V_L$ is the volume of the inductor.\cite{tinkham:1975}  For TiN, $N_0 \approx 8.7 \times 10^9$ $\mu$m$^{-3}$eV$^{-1}$.\cite{Leduc:2010}  Realizing high multiplexing densities requires a high resonator quality factor $Q_r$.  According to Eq.\ (\ref{pgCritical}), this lowers the onset of bifurcation and consequently degrades the NEP when operating below this limit.  This can be avoided by decreasing $Q_r$ however this increases the resonator bandwidth and consequently the per-pixel cost.  While it has been shown for a TiN far-infrared LEKID pixel that it is possible to operate at powers beyond the onset of bifurcation,\cite{swenson:2012} this requires a considerable increase in readout complexity which inevitably leads to increased pixel cost.

Eq.\ (\ref{fNEP}) suggests another method for improving the device NEP while operating below the onset of bifurcation.  The ratio of the reactive to dissipative response is given by  $\beta = \delta \sigma_2/\delta \sigma_1 \approx 2kT/hf$.  Historically, kinetic-inductance detectors have been designed to operate around a few GHz.  Operation in this regime is useful for providing the large per-pixel bandwidth necessary for single-event detection such as single-photon counting or phonon sensing.\cite{Mazin:2010, swenson:263511, moore:232601}  However, the fundamental bandwidth requirement for submillimeter-wave astronomical imaging is set by the telescope scan speed.  A per-pixel bandwidth of $\sim$ 100-1000 Hz is sufficient for this application.  If for multiplexing reasons we require $Q_r$ to be of order $10^5$, this sets a lower limit on the resonance frequency of $f_r \sim$ 10-100 MHz.  Designing the pixels to resonate at MHz instead of GHz frequencies increases $\beta$ by more than an order of magnitude with a proportional improvement in NEP$_{freq}$.

\begin{figure}
\begin{center}
\begin{tabular}{c}
\includegraphics[width=.65\linewidth]{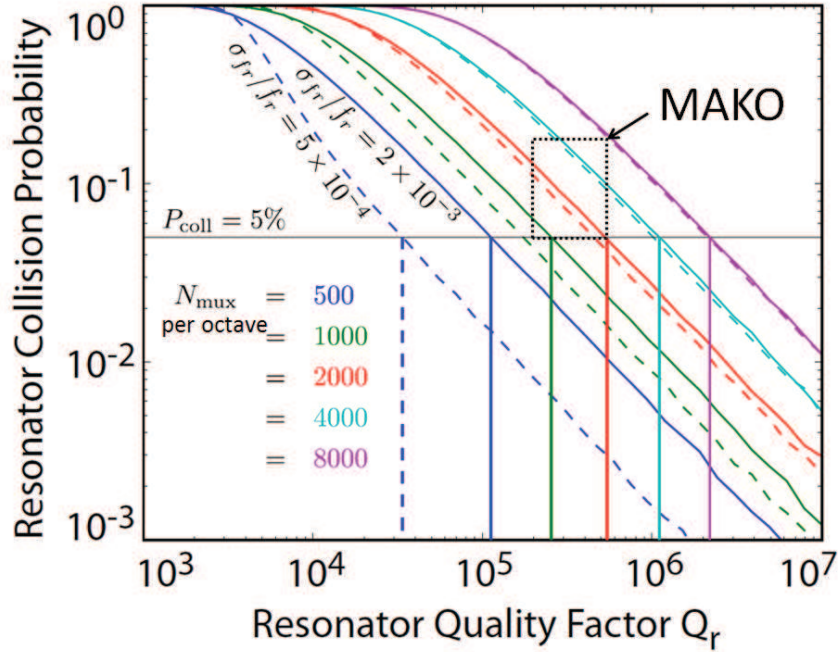}
\end{tabular}
\end{center}
\caption[example]
{
Collision probability per octave of readout bandwidth.  In this calculation, a fixed number of pixels with resonator quality factor $Q_r$ are evenly distributed throughout an octave of bandwidth.  Their positions are then perturbed randomly from their initial value $f_r$ with a gaussian distribution of width $\sigma_{f_r}/f_r$.  Two pixels are assumed to collide (ie suffer from unrecoverable electronic cross-talk) if their operating frequencies are within five linewidths.  In this case, both pixels are removed.  Typical values from previous resonator detector arrays of $\sigma_{f_r}/f_r = 2\times10^{-3}$ (solid) and $5\times10^{-4}$ (dashed) are shown.  Note that as the $Q_r$ increases, the linewidth and consequently the collision probability decrease for a fixed number of pixels per octave.  If readout cost dominates, potentially it is reasonable to accept a higher percent pixel loss if the total number of pixels per readout increases.  }
         \label{figCollisions}
\end{figure}

Operating at lower frequency provides other crucial advantages apart from improving NEP$_{freq}$.  Readout for the GHz regime typically begins by generating a frequency comb of measurement tones with a spacing corresponding to the pixel frequencies using a high resolution digital-analog converter (DAC).  These tones are located in a few hundred MHz of bandwidth around DC.  The tones are then shifted to the desired GHz band using an up-converting mixer and a separate local oscillator.  After interacting with the resonances the waveform is then shifted back down to baseband by a second mixing stage.  Subsequent digitization is accomplished using a high resolution analog-digital converter (ADC).  By operating below a few hundred MHz, the two analog mixing steps can be removed.  This reduces readout cost and complexity as well as eliminating analog mixer distortion.

The available electronic readout bandwidth is usually constrained by the sampling speed of the high resolution ($\ge 12$ bit) ADC.  Currently this provides at most a few hundred MHz of readout bandwidth.  By moving from GHz to MHz frequencies, the per-pixel bandwidth $\Delta f = f_r/Q_r$ decreases allowing a greater pixel count per fixed electronic bandwidth.  A statistical calculation of the expected pixel loss due to resonance collision for various packing densities is shown in Fig.\ \ref{figCollisions}.  Here, the fractional error in the resonator design frequency is assumed to be set by lithographic constraints.   The series of curves correspond to packing various pixel counts \emph{per octave of bandwidth}.  As the resonator $Q_r$ increases, fewer pixel collisions occur as a result of the decreased per-resonator bandwidth.  The important point here is that the number of octaves per fixed electronic bandwidth increases at lower frequencies.  For example, 250 MHz of electronic bandwidth only occupies about a quarter octave in the range 1-2 GHz whereas the same electronic bandwidth spans more than two octaves from 50 to 250 MHz.  As readout cost is roughly proportional to electronics bandwidth, increasing the multiplexing density directly lowers per-pixel cost.

\section{Measurement system}

\begin{figure}
\begin{center}
\begin{tabular}{c}
\includegraphics[width=.65\linewidth]{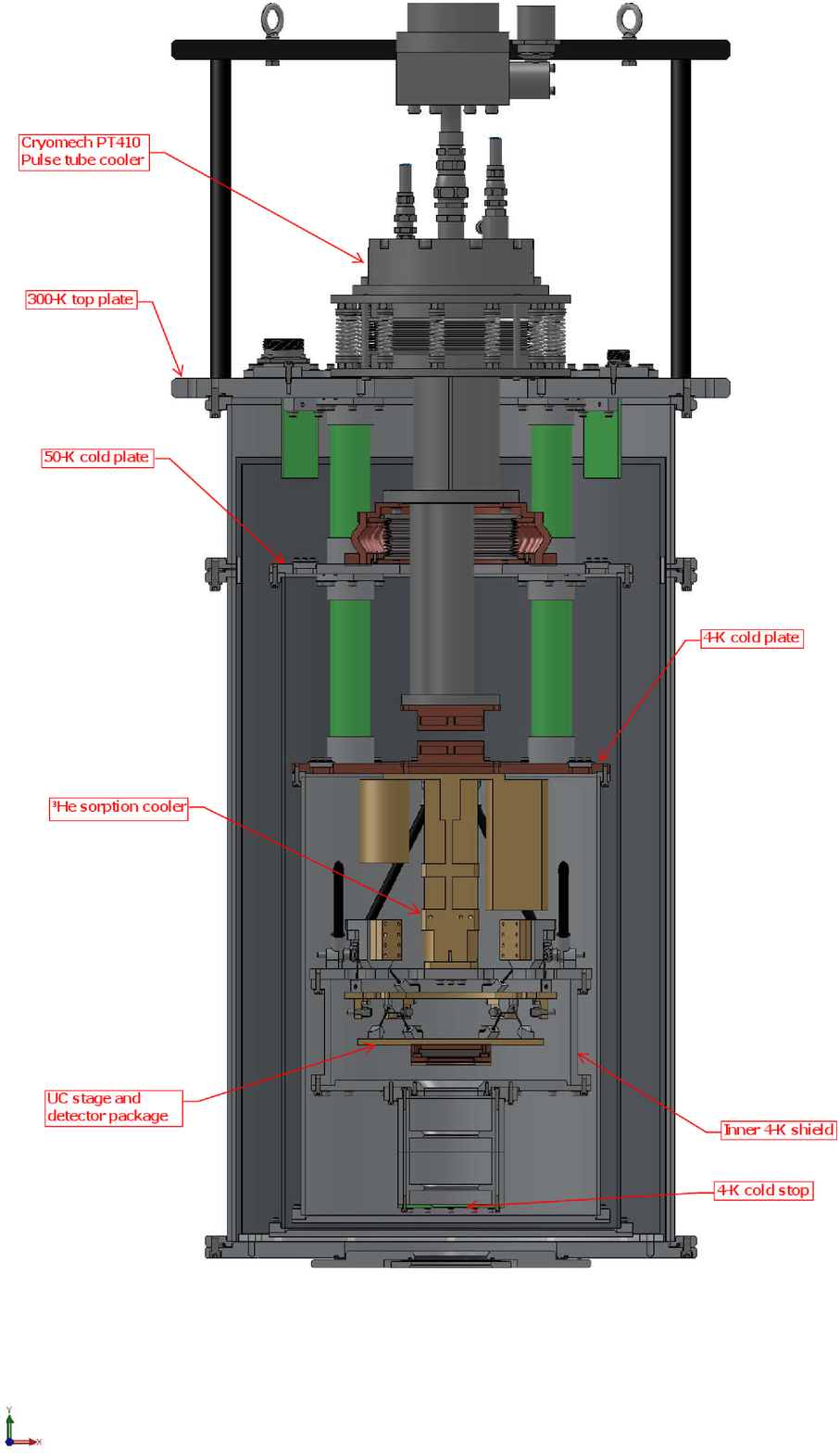}
\end{tabular}
\end{center}
\caption[example]
{ Cryomechanical and optical layout of the MAKO instrument.  This system is designed as a drop-in replacement for the SHARC-II instrument currently installed at the Caltech Submillimeter Observatory.
  }
         \label{figSystem}
\end{figure}

\subsection{Optical Design}

For simplicity, MAKO has been design to use the same relay optics
system as the existing SHARC-2 instrument \cite{dowell:73} at the
CSO. The instrument will be mounted at one Nasmyth focus of the
telescope, with two flat fold mirrors relaying the f/12.96
Cassegrain focus of the telescope to a single off-axis elliptical
mirror reducing the focal ratio to f/4.48. The
tertiary image of the far field, 45 mm in diameter, is formed 320 mm
from the elliptical mirror with a 32 mm diameter image of the primary
mirror at 178 mm. These positions lie 193 mm and 51 mm from the
cryostat reference plane, respectively. The final focus offers a 3.7
arcminute field of view, limited by the tube through the telescope
elevation bearing.

The cryostat window is composed of high-density polyethylene, 1 mm
thick with a clear aperture of 70 mm. Long-wavelength-pass filters
are mounted at the apertures of the 50 K and 4 K radiation shields
in order to reject infrared radiation from the cold stages of the
instrument. At the 50 K shield, a Z-cut crystal quartz filter of
diameter 65 mm and 2 mm thickness, with a 58 $\mu$m clear low-density
polyethylene (LDPE) coating applied to the upper surface and a
58 $\mu$m black LDPE layer adhered to the lower surface. The LDPE
layers act as an anti-reflection coating for the quartz filter,
while the black LDPE is also useful for blocking optical and
near-infrared radiation. As second, identical quartz filter is
mounted at the aperture of the 4 K radiation shield, in addition to
a 100 $\mu$m and a 300 $\mu$m metal mesh long-wavelength-pass filter,
produced by QMC Instruments.\footnote{http://www.terahertz.co.uk}
The observing band of the detectors is defined by a further metal
mesh bandpass filter, also manufactured by QMC Instruments, with a
band center at 340 $\mu$m and a 10\% bandwidth. The bandpass filter
is integrated with the detector package. The peak transmission of
the full filter system is approximately 50\%.

\subsection{Cryomechanical Design}

Since MAKO is designed to be a ``drop-in'' replacement for SHARC-2, it
was necessary that the new cryostat be mechanically compatible with
the existing hardware. Unlike SHARC-2, the MAKO cryostat uses a
Cryomech PT4-10 pulse tube mechanical cooler to provide the primary
heat uplift, whereas the current SHARC-2 system is a liquid
nitrogen/liquid helium system. The pulse tube provides a cooling
capacity of 40 W at 45 K on the first stage, used to cool a radiation
shield and optical filters as well as intercepting the conduction
load from wiring, while providing a cooling capacity of 0.9W at 4K
on the second stage. The 4 K stage of the cryostat cools a second
radiation shield with further filters and provides the heat sink for
the \hthree sorption refrigerator that cools the detector array
(described below). The vacuum jacket and radiation shields, shown in
Fig.\ \ref{figSystem} are an all-aluminium
welded design, fabricated by Precision Cryogenics
Inc.\footnote{http://www.precisioncryo.com} The outer shell of the
cryostat has a diameter of 470 mm and is 1250 mm in height, including
the open frame protecting the top plate.

The 4 K space of the cryostat is 290 mm in diameter and 420 mm in
height. In addition to the outer 4 K radiation shield, an inner
shield, composed of Aluminium 6061, encloses the sub-Kelvin stages
of the cryostat. This inner shield provides additional baffling
against stray light and the mechanical mounting of a cold stop
aperture at the image of the primary mirror. This structure is
mounted to the 4 K cold plate via a CFRP hexapod structure. Although
not intended to be thermally isolating, CFRP represents an excellent
structural material being inexpensive, light and extremely stiff.
Cooling is provided by a separate copper strap, with a reach-around
strap cooling the cold stop aperture itself rather than relying on
conduction through the aluminium structure.

The detector package is cooled by a three-stage helium sorption
fridge, design and fabricated by Chase
Cryogenics,\footnote{http://www.chasecryogenics.com} with a \hfour
cold head and two \hthree cold heads. Unlike other designs of
sorption cooler in which the \hfour stage is integrated with the
first \hthree cold head, the MAKO fridge has three separate cold
heads, although the \hfour head is used only as the condensation
point for the \hthree in the other heads. The first \hthree cold
head cools an intercept stage to approximately 350 mK, buffering the
conduction load from the 4 K stage and the superconducting coaxial
cables running to the detector package. The detector package is
cooled to ~240 mK by the second \hthree head. The intercept and
detector stages are thermally isolated with CFRP bipod trusses and
are connected to the sorption cooler with compliant copper foil
straps.

\subsection{Electronics}

State-of-the-art readout electronics for microresonator detectors currently achieve multiplexing factors around a couple hundred, with a record of 256.\cite{Mazin:2006799, swenson:84, yates:042504, bourrion:06012, mchugh:044702}  As these electronics currently cost several thousand dollars, the per-pixel cost is many tens to upwards of a hundred dollars.  Simply scaling these existing systems, the readout cost will completely dominate any system achieving tens of kilopixels.  Operating at lower frequencies results in narrow pixel linewidths which necessitates increased frequency resolution for both the individual tone generation and post-transmission signal separation compared with existing readouts.  However as already discussed, the narrow linewidths provide more than an order of magnitude gain in multiplexing density which directly translates to decreased pixel cost.  In order to realize this significant gain, we are meeting the challenge of high-resolution read out with the two electronics solutions discussed below.

Our first readout places a high priority on minimizing development time. This approach uses a dual-channel commercial digitizing card (Pentek, model 78650).  This card is equipped with a pair of DACs each connected to a circular buffer enabling looped playback of a precalculated waveform.  Dual ADCs on the card are then used to digitize the transmitted waveforms.  Rather than processing this received waveform locally, the data is streamed via a PCIe communication bus to a graphical processing unit (GPU).  The GPU performs a continuous FFT on the received data.  Selection of only the relevant FFT spectral bins and averaging then reduces the data rate before saving the results to disk.  While a field-programmable gate array (FPGA) is included on the card, it only is used for communication and signal routing and does not perform any digital-signal processing.  All FPGA functionality is preprogrammed and accessed using standard C++ libraries.  No specialized hardware development language (HDL) programming is required.

The most challenging aspect of this approach is processing the data provided by the digitizing card in real time.  The dual 12-bit ADCs each sample at a rate of 500 megasamples per second (MSPS).  The PCIe bus provides 500 MB/s per lane of bandwidth and the digitizing card utilizes 8 lanes, thus the 4 GB/s of total throughput comfortably supports the 1.5 GB/s produced by the ADCs.  Once the data has been received by the GPU, it must compute successive $N$ point FFTs on the fly.  For an ADC sampling at 500 MSPS, an FFT of at least $2^{22}$ points must be continuously calculated to achieve a $\sim$ 100 Hz resolution.  The computational cost of an FFT is $\approx 5 N\log_2(N)$ floating-point operations (FLOP) and thus the rate the gpu must support is: (ADC Sample Rate/$N$)*$5 N\log_{2}(N)$ FLOP/s.  For $N$ between $2^{20}$ to $2^{26}$, this corresponds roughly to 100 to 130 GFLOP/s to continuously support both ADCs at their full sample rate.  Modern GPUs are benchmarked to provide more than 150 GFLOP/s for single-precision FFTs providing a reasonable margin for computational overhead.

The second readout we are currently developing places a premium on algorithm efficiency and hardware cost reduction.  Currently in the initial design phase, this readout will utilize an FPGA-based ROACH platform to evaluate potential signal-processing algorithms.\cite{parsons:2006}  A variety of techniques, including multistage FFTs, a polyphase filter-bank, or a polyphase filter-bank followed by a second stage of digital down-conversion as demonstrated by McHugh et al.\cite{mchugh:044702} are being considered.  A portion of the signal processing may be performed by a host computer after an initial stage of processing reduces the information rate sufficiently for communication transfer.  The balance between FPGA and CPU/GPU processing will be optimized for cost and simplicity.  Eventually, we plan on migrating this readout to custom hardware.\cite{bourrion:06012}  Applying these optimizations should result in a readout cost reduction of around an order of magnitude.  Ultimately, the electronics will likely feature an application specific integrated chip (ASIC) in order to reduce the volume cost and readout power requirements crucial for remote-location and space-based deployment of large format instruments.

Comparison of the measurement signal to the amplifier noise affords an estimate of the highest multiplexing factor achievable on a single measurement line.  As discussed above, the maximum achievable signal power is limited by the onset of bifurcation.  For our current pixel design, $V_L = 11550 \mu$m$^3$, $T_c \sim 1.2 $K, $Q_c = 7\times10^5$, and $Q_r \sim 5\times10^4$ under 100 pW (single-polarization) of illumination from a blackbody.  Using Eq.\ (\ref{pgCritical}) and setting $E_* = E_{cond}$, the predicted bifurcation readout power is $P_g^{(c)} \approx 307$ fW = -95 dBm.  Measurements of our actual pixels however have found an onset of bifurcation at a significantly higher $P_g^{(c)} = 3.2$ pW = -85 dbm.  This is the incident microwave power from the generator.  Only a portion of this is transmitted to the amplifier; the rest is absorbed by the pixel or reflected.  For optimal coupling ($Q_i = Q_c$; $\chi_c = 1$) a quarter of $P_g^{(c)}$ is transmitted to the amplifier.  Thus at bifurcation, the signal at the amplifier is of order 800 fW.   A noise temperature of $T_a = 4$ K is typical of widely available cryogenic silicon-germanium bipolar transistor or HEMT amplifiers.  The noise spectral density of the analog electronics is then calculated to be $10\log_{10}(kT_a/P_{trans}) = -101$ dBc.  For both the readout options discussed above, the ADC is a Texas Instruments ADS5463.  The SNR for this ADC is $\sim$ 56 dBFS.  Assuming white noise and a sample rate of 500 MSPS, this corresponds to a noise spectral density of -143 dBc/Hz.  Comparing these results, around $10^4$ tones can be simultaneously multiplexed before the ADC begins to degrade the measurement signal-to-noise ratio.

\section{Conclusion}

MAKO is a 350 micron camera currently being constructed and tested at Caltech and JPL.  An on-sky demonstration of MAKO at the Caltech Submillimeter Observatory is planned for late 2012.  By emphasizing multiplexing density, fabrication simplicity, and a low per-pixel cost we aim to demonstrate that economical, large-format, far-infrared imaging arrays are readily achievable using superconducting resonator detectors.

\acknowledgments     

This research was carried out in part at the Jet Propulsion Laboratory (JPL), California Institute of Technology, under a contract with the National Aeronautics and Space Administration.  The devices used in this work were fabricated at the JPL Microdevices Laboratory.  This work was supported in part by the Keck Institute for Space Science, the Gordon and Betty Moore Foundation, and NASA grant NNX10AC83G. L. Swenson and C. McKenney acknowledge support from the Keck Institute for Space Science.  M. Hollister and L. Swenson acknowledge funding from the NASA Postdoctoral Program.  T. Mroczkowski acknowledges funding from the NASA Einstein Postdoctoral Fellowship program.

\bibliography{papersSPIE}   
\bibliographystyle{spiebib}   

\end{document}